\documentstyle[sprocl]{article}

\input epsf

\def\Journal#1#2#3#4{{#1} {\bf #2}, #3 (#4)}

\def\NPB{{\em Nucl. Phys.} B}

\def\PRD{{\em Phys. Rev.} D}

\def\beq{\begin{equation}}
\def\eeq{\end{equation}}

\begin{document}

\title{HIGH ENERGY INTERACTION WITH THE NUCLEUS IN THE
 PERTURBATIVE QCD WITH  $N_c\rightarrow\infty$}

\author{M.A.BRAUN}

\address{ Department of High Energy physics, University
of S.Petersburg,\\198904 S.Petersburg, Russia\\
E-mai: braun1@snoopy.phys.spbu.ru}

\maketitle\abstracts
{The BFKL fan diagram equation
for the scattering on the
nucleus  is solved numerically with the eikonalized initial condition and
for a realistic nuclear density. The gluon density has a soliton-like
form in the $\log q - y$ space. Inclusive cross-sections for jet
production in hA and AB collisions are calculated.}

%%%%%%%%%%%%%%%%%%%%%%%%%%%%%%%%%%%%%%%%%%%%%%%%%%%%%%%%%%%%%%%%%%%
%%%%%%%%%%%%%%%%%%%%%%%%%%%%%%%%%%%%%%%%%%%%%%%%%%%%%%%%%%%%%%%%%%%
\section{Introduction}
In the  colour dipole approach
with
$N_c\rightarrow\infty$ ~\cite{mu1,mu2}
the interaction
with a heavy nucleus is exactly described by  a sum of fan diagrams
constructed 
of BFKL pomerons, each of them splitting into two ~\cite{br1,ko1}. 
The equation for this sum has been
constructed in ~\cite{ko1,bal,br2}.
Let the forward scattering amplitude on the nucleus at fixed impact
parameter $b$ be
\beq
{\cal A}(y,b)=2is\int d^2r\rho(r)\Phi(y,b,r).
\eeq
where $\rho$ is the colour density of the projectile and $\Phi$
represents the sum of all BFKL fan diagrams attached to it.
Then function $\phi=\Phi/(2\pi r^2)$ in the momentum space satisfies
a non-linear evolution equation in the rapidity
\beq
\left(\partial/\partial y+H\right)\phi(y,q,b)=-\phi^2(y,q,b),
\eeq
where $y=Y/\bar{\alpha}$ with $\bar{\alpha}=\alpha_sN_c/\pi$ is a
rescaled rapidity and $H=\ln r^2 +\ln q^2+const$ is the forward BFKL
Hamiltonian. Eq. (2) should be solved with an initial condition which
follows from  the Glauber form of $\Phi$ at zero rapidity:
\beq
\Phi(0,r,b)=1-\exp\Big(-8\pi^2AT(b)\int d^2r'G(0,r,r')\rho_N(r')\Big).
\eeq
Here $G$ is the forward BFKL Green function, $T$ is the nucleus
profile function and $\rho_N$ is the colour
density of the nucleon from the target nucleus.
Eq. (2) was
studied perturbatively in ~\cite{ko2}, by asymptotic estmates in
~\cite{lev}
 and finally
solved numerically in ~\cite{br2} under the simplifying assumptions that
the nucleus has a finite radius and that  multiple Glauber interactions
in (3) can be neglected, which is justified if $A^{1/3}<<N_c$.

At small $q$ the solution of (2) behaves like $ const-\log q$
and consequently $\Phi(y,r,b) \to 1$ at $y\to\infty$ and fixed $r$
and $b$. This corresponds to the saturation at high rapidities
of the cross-section
for the scattering of a colour dipole on the nucleus to its unitary
black disk limit. However this does not imply that the structure
function of the nucleus saturates, since the number of colour dipoles
generated by the virtual photon is unlimited. In fact, as argued in
~\cite{ko2,lev} and confirmed by numerical calculations in
~\cite{br2}, it grows with $y$.
 The
gluon density of the nucleus in the combined momentum-impact parameter
space 
\beq
\partial xG(x,q,b)/\partial^2q\partial^2b=\partial^2\phi/
(2\pi\bar{\alpha}(\partial \ln q)^2)
\equiv h/(2\pi\bar{\alpha}),\ \ y=-\ln x
\eeq
was found to be
a soliton wave in $y-\ln q$ space moving
towards higher rapidities with a constant velocity and preserving its
nearly Gaussian shape. A fit to numerical data with  $\xi=\ln q$ gives
\beq
h(y,q,b)=h_0e^{-a(\xi-\xi_0(y,b))^2},\ \
\xi_0=-3.11+(2/3)\ln B+\Delta_0 y,
\eeq
where
 $\Delta_0=2.3\bar{\alpha}$,
$h_0\simeq 0.3$ and $a\simeq 0.3$ are universal.
Dimensionless parameter
$B=\pi\alpha_s^2AT(b)R_N^2$
where $R_N$ is the nucleon radius.
Actually the asymptotics of $h$ at high $|\xi|$ is not Gaussian
but exponential: at $\xi\to\infty$ $h\sim\exp(-1.3\xi)$ and at
$\xi\to -\infty$ $h\sim\exp 2\xi$.

Here we first report on the improved numerical solution of Eq. (2)
which takes into account multiple Glauber collisions in (3) and
a realistic form of nucleus.
Next we consider
simplest of the production processes: the
single  inclusive jet production in hA and AB collisions and
double inclusive jet production in hA collisions.
%%%%%%%%%%%%%%%%%%%%%%%%%%%%%%%%%%%%%%%%%%%%%%%%%%%%%%%%%%%%%%%%%%%%%%
\section{Improved nucleus structure function and gluon densities}
Our first improvement consists in taking into account the full
Glauber series in the initial value (3). 
Since at large $b$  and consequently small $T(b)$ the contribution
of the multiple scattering is neglegible, we used the same simplified
profile function corresponding to a finite nucleus as in ~\cite{br2}.
It was noted in ~\cite{br2} that the solution of (2) very quickly forgets its
initial form as $y$ grows, so that one expects to notice the influence
of the glauberization of the initial function only at the initial
stage of the evolution. This is confirmed by  numerical calculations.
The gluon densities obtained from the full initial function (3)
and  from its
single rescattering term only differ at small $y$ and $q$, where the
former tends to zero and the latter tends to a constant. This
difference turns out to be completely
washed out already at $y=1$. For $y\geq 1$ the two gluon densities
are identical for all values of $q$.
So our first conclusion is that
the Glauber rescattering at small $y$ has practically no influence
on the behaviour of the interaction with the nucleus at high
rapidities.

%%%%%%%%%%%%%%%%%%%%%%%%%%%%%%%%%%%%%%%%%%%%%%%%%%%%%%%%%%%%%%%%%
Next we studied the contribution of very peripheric collisions for
realistic nuclei with exponentially falling profile function. The
interest in this effect is due to the fact that this contribution
is not damped by the non-linear term and grows exponentially with $y$
until $\phi\sim 1$. If $T(b)\sim\exp(-b/R_A)$ then this happens at
$b\sim b_0=R_A\Delta y$ where $\Delta =4\ln 2$ is the BFKL intercept.
The total cross-section for the scattering of a colour dipole on
the nucleus at $b>b_0$ and $y>>1$  then grows
linearly with $y$. Corespondingly one  expects that the
structure function will grow with $y$ faster than linearly.
These expectations are also confirmed by our numerical
calculations. At large values of $1/x=\exp Y$ ($\sim 10^{15}\div 10^{20}$)
the found structure functions of Pb 
with a constant density inside a finite sphere (CD) and with
the Woods-Saxon nuclear density (WS)  can be fitted by 
\[ F_2^{CD}\simeq Q^2\,(332 y-73 \ln Q^2-893),\]
\[ F_2^{WS}\simeq Q^2\, (12.4 y-2.58 \ln Q^2-29.0)^2, \]
where $y=\bar{\alpha}Y$ and $Q^2$ is in (GeV/c)$^2$.
Thus, to a good approximation, the structure function of a
realistic nucleus grows with $1/x$ as $\ln^2 (1/x)$.
%%%%%%%%%%%%%%%%%%%%%%%%%%%%%%%%%%%%%%%%%%%%%%%%%%%%%%%%%%%%%%%%%%%%

Finally a few words about our definition of the gluon density in
relation to  the saturation phenomenon as discussed by A.Mueller ~\cite{mu3}.
We define the gluon density as essentially
the quantity to be integrated with the
quark loop to obtain the virtual photon scattering cross-section
and structure function in the low $x$ kinematical
region. This definition is appropriate for this
kinematical region, without strong ordering in the transverse
momentum, and cannot be directly applied to the large $x$
region where such ordering takes place.
This circumstance has to be taken into account when comparing our
gluon density with differently defined ones. In particular
the gluon density introduced in ~\cite{mu3} via the interaction of a "gluonic
current" with a nucleus is a totally different quantity,
more appropriate for the large $x$ region. One can show that
A.Mueller's density can be expressed in terms of our density (4) and
that its saturation properties  obtained in ~\cite{mu3} follow from the found
solitonic form of (4) ~\cite{br3}.

%%%%%%%%%%%%%%%%%%%%%%%%%%%%%%%%%%%%%%%%%%%%%%%%%%%%%%%%%%%%%%%%% 
\section{Single jet inclusive production}
%%%%%%%%%%%%%%%%%%%%%%%%%%%%%%%%%%%%%%%
Single jet production inclusive cross-sections in hA collisions
are obtained from the imaginary part of forward scattering
amplitude $\cal A$  (Eq. (1)) by
a substitution of one of the BFKL Green functions in the fan diagrams
\beq
G(Y)\rightarrow G(Y-y)V_k(r)G(y),\ \
V_k(r)=(4N_c\alpha_s/k^2)\stackrel{\leftarrow}{\Delta} e^{ikr}
\stackrel{\rightarrow}{\Delta}
\eeq
(the arrows shows the direction of differentiation).
Due to the AGK rules contribution of all such substitutions
below the uppermost spliiting point cancel so that we are left with
\beq
d\sigma/dy d^2kd^2b=2\langle\rho G(Y-y)V_k\Phi(y)\rangle
=(16\pi^2\bar{\alpha}/k^2)
\langle\rho [G(Y-y)\stackrel{\leftarrow}{\Delta}]e^{ikr}h(y)\rangle ,
\eeq
where $\rho$ and $\Phi$ are from Eq. (1), $\langle...\rangle$
means integrating over the gluon relative transverse coordinates or
momenta and  the relation between $\Phi$ and $\phi$ and the explicit form
of $V_k$ have been used to obtain the second equality.

Using the form of $h$ obtained numerically we find jet multiplities at large
$y$ and $Y-y$
\beq
\mu_A(Y,y,k)=(c/k^2)A^{2/9}
e^{\Delta Y-\epsilon y}/\sqrt{y(Y-y)}
\eeq
where $\Delta=4\ln2$, $\epsilon=0.39$, $c$ is a known numerical
constant and all rapidities are scaled with $\bar \alpha$.
The inclusive cross-section integrated over $k$ evidently  diverges at
low $k^2$. Physically relevant results correspond to jets with  not
too small transverse momentum $k>k_{min}$.

The $A$ dependence is different both from the
eikonal and local pomeron fan diagrams predictions. The $k$ dependence
is  $1/k^2$ up to  $\ln k\sim \Delta y$ in contrast
to the $hh$ scattering where it is cut by the damping factor
$\exp(-c\ln^2k/y)$. In this way we observe the "Cronin effect":
the distribution for the nucleus target is flatter than for the hadronic one.

These results are easily generalized to nucleus-nucleus scattering.
As was shown in ~\cite{ama}, for them the single inclusive cross-section is
given by the sum of fan diagrams connecting the produced jet with the
both participant nuclei.  So at fixed impact parameters of the nuclei
$b_A$ and $b_B$  the inclusive cross-section is given by an obvious
generalization of Eq. (7):
\beq
d\sigma/dy d^2kd^2b_Ad^2b_B
=2\langle\Phi_A(Y-y)V_k\Phi(y)\rangle=(32\pi^3\bar{\alpha}/
k^2)\langle h_A(Y-y)e^{ikr}h_B(y)\rangle
\eeq

Using our results for the gluon densities, for $\ln k<<\xi_0$,
$y\sim Y/2$ and identical nuclei, we obtain rough estimates
$\sim (1/k^2)A^{4/9}e^{\Delta_1Y}$
for the inclusive cross-section at a given $k$ and
$\sim A^{2/3}e^{\frac{3}{2}\Delta_1Y}$
for the cross-section integrated over $k>k_{min}$
with $\Delta_1=2.4$. Note that from the latter estimate it follows that
the multiplicity density at a given $y$ behaves as $A^{4/3}$, that is,
similarly to the eikonal result.

These estimates are supported by our numerical calculations.
For Pb-Pb collisons at center rapidity with $\alpha=0.16$
we obtained values for the multiplicity which are well described
by a fit
\[\mu(0)=0.092\, e^{0.59 Y}\,(1+35.e^{-0.15 Y}).\] 
Note that with $\alpha=0.16$ the BFKL pomeron
intercept is $\Delta=0.42$, so that the multplicities  grow
faster than the BFKL pomeron.
The $A$ dependence of $\mu(0)$ at a given $Y$ is almost perfectly
represented by a power factor
$\mu(0)\sim A^{\beta(Y)}$,
the power $\beta(Y)$ slowly rising with $Y$ from
1.25 at $Y=10\div 20$ to 1.29 at $Y=40$. So it seems that the
difference in the $A$ dependence with the eikonal prediction
($\beta=4/3$) is gradually disappearing at high enough rapidities.

The form of the distribution in $y$ turns out to be
practically independent of the
atomic number.  With the growth of
energy the distribution becomes more strongly peaked at the center.

\section{Double inclusive cross-sections}
For the nucleus-nucleus collisions the double inclusive cross-sections
cannot be expressed via sums of fan diagrams for the participant nuclei, but
require summation of diagrams of a more complicated structure ~\cite{ama}.
For this
reason we restrict ourselves to the case of hA scattering.
The double inclusive cross-section is obtained from the imaginary part of the
amplitude (1) by two substitutions (6) for the two produced jets with
rapidities $y_{1,2}$ and transverse momenta $k_{1,2}$. The AGK rules
leave two contributions corresponding to the two substitutions
 either in the uppermost pomeron or  in the two pomerons immediately
 after the first splitting. These two contributions rise with the
 overall rapidity $Y$ as
 $e^{\Delta Y}$  and $e^{2\Delta Y}$ respectively, so that the second one
is dominant at high $Y$.

Study of this contribution shows that the multiplicity ratio
\beq
R=\frac{\mu(y_1,k_1;y_2,k_2)}{\mu(y_1,k_1)\mu(k_2,y_2)}=
c\sqrt{\frac{(Y-y_1)(Y-y_2)}{2Y-y_1-y_2}},
\eeq
where $c$ is a known constant proportional to $\alpha_s$,
does not depend on $k_{1,2}$,
grows with $Y$ at center rapidity and stays constant in the
diffractive regions. With $\alpha_s= 0.2$ at $y=Y/2$ we find
$R=0.19\sqrt{Y}$ so that correlations change from negative to positive at
$Y\sim 25$.

\section{Conclusions}
Improved numerical calculations confirm the validity of approximations
made in solving Eq. (2) in ~\cite{br2}.
The nuclear structure function seems to grow with rapidity also
for a realistic nucleus with a profile function exponentially falling
at large impact parameters, the growth being faster than for a finite
nucleus.

The inclusive jet production cross-sections grow rapidly with energy as
the pomeron itself or even faster (for AA collisions). Their $A$-
dependence results close to the eikonal predictions in evident
contrast to the earlier ones made for fan diagrams with a local pomeron.

\section*{Acknowledgments}
The author greatly benefited by fruitful discussions with
B.Kopeliovich, Yu. Kovchegov, E.Levin and A.Mueller,
to whom he expresses his deep gratitude. This work has been
supported by NATO grant PST.CLG.976799.

\end{document}